\documentclass[fleqn,usenatbib,usedcolumn]{mnras}
\usepackage[british]{babel}            
\usepackage{newtxtext}                  
\usepackage[slantedGreek]{newtxmath}  
\usepackage[T1]{fontenc}  
\usepackage{amsmath}
\usepackage{epsfig} 
\usepackage{graphicx}	
\usepackage{amssymb}	

\newcommand{\be}{\begin{equation}}
\newcommand{\e}{\end{equation}}
\newcommand{\bear}{\begin{eqnarray}}
\newcommand{\ear}{\end{eqnarray}}

\newcommand{\del}{\partial}

\def\aj{AJ}
\def\apj{ApJ}
\def\apjs{ApJS}
\def\jcap{JCAP}
\def\mnras{MNRAS}
\def\aap{A\&A}
\def\prd{Physical Review D}
\def\prl{Physical Review Letters}
\def\nat{Nature}      
\def\apjs{ApJS}
\def\apjl{ApJ Letters}

\def\qjras{QJRAS}

\title[Entropy, expansion and acceleration ] {Does information entropy
  play a role in the expansion and acceleration of the Universe?}
    
\author[Pandey, B.] {Biswajit Pandey\thanks{E-mail:
    biswap@visva-bharati.ac.in}
  \\ Department of Physics, Visva-Bharati University, Santiniketan,
  Birbhum, 731235, India\\ }
   
 \date{\today}

 \pubyear{2017}
  
\begin{document}
\label{firstpage}
\pagerange{\pageref{firstpage}--\pageref{lastpage}}      
\maketitle
       
\begin{abstract}

We propose an interpretation of the expansion and acceleration of the
Universe from an information theoretic perspective. We obtain the time
evolution of the configuration entropy of the mass distribution in a
static Universe and show that the process of gravitational instability
leads to a rapid dissipation of configuration entropy during the
growth of the density fluctuations making such a Universe entropically
unfavourable. We find that in an expanding Universe, the configuration
entropy rate is governed by the expansion rate of the Universe and the
growth rate of density fluctuations. The configuration entropy rate
becomes smaller but still remains negative in a matter dominated
Universe and eventually becomes zero at some future time in a
$\Lambda$ dominated Universe. The configuration entropy may have a
connection to the dark energy and possibly plays a driving role in the
current accelerating expansion of the Universe leading the Universe to
its maximum entropy configuration.
\end{abstract}


\section{Introduction}
The expanding Universe \citep{hubble} and its current accelerating
expansion \citep{riess,perlmutter1} are the two milestone discoveries
in observational cosmology. In the current standard model of
cosmology, the accelerating expansion of the Universe is explained by
an all permeating hypothetical form of energy known as the dark
energy. Independent observations from high redshift supernova
\citep{riess,perlmutter1}, cosmic microwave background radiation
\citep{sherwin,hinshaw,planck16a}, gravitational lensing
\citep{sarbu,weinberg, cao} and large scale structures
\citep{perlmutter2, blanchard, pavlov} suggest that the dark energy
constitutes almost $\sim 70\%$ of the total energy density of the
present day Universe. However the exact nature of the dark energy is
still unknown and remains matters of speculation. The $\Lambda$CDM
model is currently the most successful model in explaining a wide
range of cosmological observations. In the $\Lambda$CDM model, one may
choose to identify the cosmological constant as the dark
energy. However this leads to a discrepancy by an embarrassing $120$
orders of magnitude between the tiny observed value of the
cosmological constant and its large theoretically predicted value.
Various alternatives to dark energy for explaining the observed
accelerated expansion has been proposed since its
inception. \citet{buchert2k} used the backreaction mechanism resulting
from the non-commutativity of time evolution and spatial averaging in
a perturbed Universe to explain the accelerating expansion. The
presence of a large local void can also mimic an apparent acceleration
of expansion \citep{tomita01, hunt}. The entropic force arising from
the information storage on the horizon surface screen can also produce
an acceleration \citep{easson}. Some more recent works suggest that the
non-zero vacuum energy density of the Universe is related to the
finite amount of accessible information stored in the spacetime
\citep{paddy, paddyhamsa}. At present, understanding the dark energy and the
cosmic acceleration remains one of the most profound puzzles in
cosmology and science in general.

In the present work, we consider the information entropy of the
Universe together with the other sources of entropy generation and
apply the maximum entropy production principle to understand the
expansion and acceleration of the Universe. The present entropy budget
of the Universe is made up of several components. One can calculate
the contributions from the CMB photons, stellar photons, neutrinos and
gravitons to the total entropy \citep{egan}. The entropy of these
relativistic particles do not change during the expansion of the
Universe. The entropy of the baryonic matters in stars, interstellar
medium (ISM) and the intergalactic medium (IGM) can be estimated from
the Sackur-Tetrode equation \citep{basu,egan} which gives the entropy
per baryon. The entropy in the baryonic component decreases with time
as it becomes more structured during the evolution of the
Universe. The growth of the Stellar Black Holes (SBH) and the
Supermassive Black Holes (SMBH) are also known as two efficient
sources of entropy in the Universe. The entropy of a Schwarzschild
black hole is well known \citep{bekenstein,hawking} which can be
integrated over the SBH and SMBH mass functions to estimate their
contributions to the total entropy of the Universe
\citep{penrose,frampton,egan}. Further, most of the mass in the
Universe is in the form of dark matter. Calculating the entropy of the
dark matter is tricky as its exact nature is unknown. However, one can
calculate it assuming that the dark matter is weakly interacting
massive particle (WIMP) \citep{egan}. Besides these contributions, the
total entropy within the Cosmic Event Horizon (CEH) must also include
an additional contribution from CEH itself which can be calculated
using the formula provided by \citet{gibbons}. The entropy of the
horizon largely dominates the other sources of entropy in the Universe
\citep{egan}.

Finally, the mass distribution in the present Universe is highly
organized into complex hierarchical patterns. The galaxies are found
to be distributed in a complex network of clusters and filaments
surrounded by voids. This complex network is often referred as the
cosmic web \citep{bond96}. The cosmic web emerges naturally by the
process of gravitational instability which amplifies the tiny density
fluctuations seeded in the early Universe. We argue that one should
also take into consideration the change in the entropy of the Universe
associated with its evolution from a nearly smooth distribution to a
highly organized complex structure. We define the configuration
entropy of the mass distribution based on the idea of information
entropy \citep{shannon48} and study its time evolution in the linear
regime using the linear perturbation theory. If the second law of
thermodynamics applies to our Universe as a whole then the sum of all
the entropies in the Universe must increase with time since the big
bang. We study how the time evolution of the configuration entropy
affects the evolution of the total entropy of the Universe which may
in turn influence the dynamics of the Universe on large scales.

The connection between the cosmic accelerated expansion and the second
law of thermodynamics has been studied extensively \citep{radicella,
  pavon1, mimoso, pavon2, ferreira}. The accelerated expansion of the
Universe is consistent with the second law of thermodynamics and it
could have been anticipated on thermodynamic grounds before its
discovery from the observations of high redshift supernova
\citep{radicella, pavon2}. Based on the Hubble expansion history,
\citet{pavon2} suggested that the entropy of the Universe tends to
some maximum value. Interestingly different cosmological models such
as nonsingular bouncing Universes, modified gravity theories and the
Universes whose expansion is dominated by matter or by phantom fields
do not tend to a state of maximum entropy whereas the $\Lambda$CDM
model does \citep{radicella, mimoso, ferreira}.

The information entropy has been used earlier in cosmology to study
various issues like homogeneity \citep{hosoya,pandey1}, isotropy
\citep{pandey2}, complexity \citep{vazza}, bias and non-Gaussianity
\citep{pandey3}, information entanglement \citep{czinner}, dark energy
from entanglement \citep{capoluongo1,capoluongo2} and the large scale
environmental dependence of galaxy properties \citep{pandey4}. In the
present work, we use information entropy to understand the expansion
and acceleration of the Universe.

A brief outline of the paper follows. In section 2 we describe the
configuration entropy and its time evolution in a static and an
expanding Universe. We discuss the possible roles of the configuration
entropy in the observed expansion and acceleration of the Universe and
present our conclusions in section 3.

\section{CONFIGURATION ENTROPY AND ITS TIME EVOLUTION}

The continuity equation in a fluid is,  
\begin{eqnarray}
\frac{\del \rho}{\del t} + \nabla \cdot (\rho \vec{v})& = & 0
\label{eq:one}
\end{eqnarray}
where $\rho$ is the fluid density and $\vec{v}$ is the flow velocity.
We define the configuration entropy of the fluid as,
\begin{eqnarray}
S_{c}(t) & = & -\int\int\int \rho \log \rho\, dx dy dz
\label{eq:two}
\end{eqnarray}
where the entire space occupied by the fluid is divided into a finite
number of elements with volume $dV = dx dy dz$.

Multiplying \autoref{eq:one} by $(1+\log \rho)$ and integrating over
the whole space we get,
\begin{eqnarray}
\frac{dS_{c}}{dt} - \int \nabla \cdot (\rho \log \rho \, \vec{v}) \, dV - \int \rho \, \nabla \cdot \vec{v} \, dV & = & 0 
\label{eq:three}
\end{eqnarray}
We consider a sufficiently large volume of the Universe. The second
term in the \autoref{eq:three} can be expressed as a surface integral
which vanishes at the boundary. So we finally have,
\begin{eqnarray}
\frac{dS_{c}}{dt} & = & \int \rho \, \nabla \cdot \vec{v} \, dV 
\label{eq:four}
\end{eqnarray}

A similar relation has been obtained by \citet{liang} in the context
of information transfer between dynamical system components.

\subsection{CONFIGURATION ENTROPY IN A STATIC UNIVERSE}
We consider a self-gravitating fluid with small fluctuations in its
density, pressure and potential in a static Universe. This leads to
the well known Jeans instability in a static Universe. In this case
the linearized continuity equation becomes,
\begin{eqnarray}
\nabla \cdot \vec{v}=-\frac{\del \delta}{\del t}
\label{eq:five}
\end{eqnarray}
where $\delta(\vec{r},t)=\frac{\rho(\vec{r},t)-\bar{\rho}}{\bar{\rho}}$ is
the density contrast and $\bar{\rho}$ is the mean density.

Substituting $\nabla \cdot \vec{v}$ in \autoref{eq:four} gives,
\begin{eqnarray}
\frac{dS_{c}}{dt} & = & - \int \bar{\rho} (1+\delta)\,\frac{\del \delta}{\del t} \, dV 
\label{eq:six}
\end{eqnarray}
As of now we have only used the equation of continuity which ensures
mass conservation during the fluid flow. Combining the Euler's
equation and Poisson's equation with the continuity equation yields
the equation governing the growth of density perturbations
$\delta(\vec{r},t)$ in a static Universe. The solution of this
equation tells us that the density perturbations larger than the Jeans
length will grow exponentially as
$\delta(\vec{r},t)=\delta(\vec{r}) \exp({\sqrt{4 \pi G
    \bar{\rho}}\,t})$.  Substituting this in \autoref{eq:six} and
simplifying we get,
\begin{eqnarray}
\frac{dS_{c}}{dt} & = & - \bar{\rho}  \sqrt{4 \pi G \bar{\rho}} \,\exp({2\,\sqrt{4 \pi G \bar{\rho}}\,t})\int \delta^{2}(\vec{r}) \, dV 
\label{eq:seven}
\end{eqnarray}
 The integral in \autoref{eq:seven} is related to the variance of the
 fluctuations and is a positive quantity. This suggests that the
 configuration entropy rate of a self-gravitating fluid in a static
 Universe is negative and its magnitude blows up exponentially with
 time.

\subsection{CONFIGURATION ENTROPY IN AN EXPANDING UNIVERSE}

Now we consider the growth of the density perturbations in a self
gravitating fluid in an expanding Universe. The continuity equation in
the comoving co-ordinate is,
\begin{eqnarray}
\frac{\del \rho}{\del t} + 3\, \frac{\dot{a}}{a} \rho+\frac{1}{a}\,
\nabla \cdot (\rho \vec{v})& = & 0
\label{eq:eight}
\end{eqnarray}
where $a$ is the cosmological scale factor and $\vec{v}$ is the
peculiar velocity.

Multiplying \autoref{eq:eight} by $(1+\log\rho)$ and integrating over
the whole space gives,
\begin{eqnarray}
\frac{dS_{c}(t)}{dt} + 3 H(t)\,S_{c}(t) - \frac{1}{a} \int \rho \, (3 \dot{a}+\nabla \cdot \vec{v}) \, dV & = & 0 
\label{eq:nine}
\end{eqnarray}
where $H(t)=\frac{\dot{a}}{a}$ is the Hubble parameter. We express the
third term in the left hand side of \autoref{eq:nine} as some function
of time $F(t)=\frac{1}{a} \int \rho \, (3 \dot{a}+\nabla \cdot
\vec{v}) \, dV$.  One may note that $F(t)$ can be positive, negative
or zero for the different parts of the fluid volume. $F(t)$ will be
zero when the divergence of the peculiar flow is cancelled by the
divergence of the Hubble flow, positive when the Hubble flow dominates
over the peculiar flow and negative otherwise.

We now have an ordinary differential equation,
\begin{eqnarray}
\frac{dS_{c}(t)}{dt} + 3 H(t)\,S_{c}(t) - F(t) & = & 0
\label{eq:ten}
\end{eqnarray}
The solution of \autoref{eq:ten} gives the evolution of the
configuration entropy in an expanding Universe which is given by,
\begin{eqnarray}
S_{c}(t)=\exp(-3\int_{t_{0}}^{t} H(t^{\prime})
\,dt^{\prime})\Big[S_{c}(t_{0}) \nonumber \\+\int_{t_{0}}^{t} F(t^{\prime\prime})
\,\exp(3\int_{t_{0}}^{t^{\prime\prime}} H(t^{\prime})
  \,dt^{\prime}) dt^{\prime \prime}\Big]
\label{eq:eleven}
\end{eqnarray}
where $S_{c}(t_{0})$ is the configuration entropy at time
$t_{0}$. Noticeably, the magnitude of the entropy rate becomes smaller
in an expanding Universe as compared to a static Universe and it would
depend on the form of the function $F(t)$. We can express $F(t)$ as,
\begin{eqnarray}
F(t)& = & 3 M H(t)+\frac{1}{a}\int \rho(\vec{x},t) \, \nabla \cdot \vec{v}
\, dV \nonumber \\ & = & F_{1}(t)+F_{2}(t)
\label{eq:twelve}
\end{eqnarray}
where $\vec{x}$ is the comoving co-ordinate and $M=\int
\rho(\vec{x},t) \, dV=\int \bar{\rho}(1+\delta(\vec{x},t))\, dV$ is
the total mass enclosed inside the comoving volume $V$.

One can simplify $F(t)$ further using the linear perturbation
theory. The linearized continuity equation in the comoving co-ordinate
becomes,
\begin{eqnarray}
\nabla \cdot \vec{v}=- a \frac{\del \delta}{\del t}
\label{eq:thirteen}
\end{eqnarray}
In the linear regime, the shape of the density fluctuations remain
frozen in comoving co-ordinates and their amplitude grows as
$\delta(\vec{x},t)=D(t) \delta(\vec{x})$ where $D(t)$ is the growing
mode of density perturbations. The second term in \autoref{eq:twelve}
can be expressed as,
\begin{eqnarray}
F_{2}(t) & = & - \bar{\rho} \dot{D}(t) D(t) \int
\delta^{2}(\vec{x})\,dV \nonumber\\ & = &- \bar{\rho} H(t) f D^{2}(t)
\int \delta^{2}(\vec{x})\,dV
\label{eq:fourteen}
\end{eqnarray}
where $f=\frac{dlnD}{dlna}$ is the logarithmic derivative of the
growing mode with respect to the scale factor. 

\subsection{CONFIGURATION ENTROPY RATE IN A MATTER DOMINATED UNIVERSE}

We use the linear perturbation theory to calculate the configuration
entropy rate in the linear regime in a matter dominated expanding
Universe. The configuration entropy rate depends on the form of the
function $F(t)$ in \autoref{eq:twelve}. In case of a matter dominated
Universe, we have $H(t) \propto \frac{1}{t}$, $D(t) \propto a(t)$ and
$a(t) \propto t^{\frac{2}{3}}$. Consequently $F_{1}(t)$ will scale as
$t^{-1}$ and $F_{2}(t)$ will scale as $t^{\frac{1}{3}}$. Since
$F_{2}(t)$ is negative so $F(t)$ will become negative after large $t$.

This suggests that the configuration entropy rate decreases
significantly in a matter dominated Universe but still remains
negative after a long time. This is related to the ongoing growth of
perturbations on all scales in a matter dominated Universe.

\subsection{CONFIGURATION ENTROPY RATE IN A $\Lambda$ DOMINATED UNIVERSE}

Now we analyze the situation in a $\Lambda$ dominated Universe. In a
$\Lambda$ dominated Universe, $H(t)=constant$, $D(t)=constant$ and
$a(t) \propto \exp(H t)$. Thus $F_{1}(t)$ becomes a constant and
$F_{2}(t)=0$. As a result the first term in \autoref{eq:eleven} is
exponentially damped and the second term converges to a constant
value. When $S_{c}(t)$ reaches a constant value, the configuration
entropy rate $\frac{dS_{c}}{dt}=0$.

This suggests that the configuration entropy converges to a constant
value in the linear regime in a $\Lambda$ dominated Universe and there
is no longer any increase or decrease in the configuration entropy on
large scales. The density perturbations stop growing on large scales
in an $\Lambda$ dominated Universe leading to $\frac{dS_{c}}{dt}=0$
which maximizes the entropy production.

\section{DISCUSSION AND CONJECTURE}

We find that the configuration entropy decreases when the Universe
self-organizes itself into complex patterns from a nearly homogeneous
and isotropic fluid with tiny fluctuations seeded in the early
Universe. If the second law of thermodynamics applies to the Universe
as a whole then the total entropy of the Universe must increase with
time. The total entropy of the Universe is distributed into several
components. Let us write the total entropy inside a significantly
large comoving volume in the Universe as
$S_{T}(t)=S_{other}(t)+S_{c}(t)$ where $S_{other}(t)$ is the sum of
all other entropies in that volume except the configuration entropy
$S_{c}(t)$. The volume is bounded by a closed comoving surface and the
large scale homogeneity and isotropy allows us to treat this volume as
an isolated system as there are no net flows across the neighboring
volumes. Consequently, we can claim that
$\frac{dS_{T}(t)}{dt}=\frac{dS_{other}(t)}{dt}+\frac{dS_{c}(t)}{dt}>0$
inside that volume. This applies to all such large volumes and the
Universe as a whole. The \autoref{eq:seven} shows that
$\frac{dS_{c}(t)}{dt}<0$ and the magnitude of the dissipation rate of
the configuration entropy grows exponentially with time when the
structure formation continues in a static Universe. This implies that
in the long run, the negative configuration entropy rate will
eventually dominate the total entropy rate leading to a negative total
entropy rate $\frac{dS_{T}(t)}{dt}<0$ in that volume. This is
particularly important when the other entropy generation processes are
not as efficient as the dissipation of the configuration entropy in
that volume. None of the known entropy generation mechanisms in the
Universe can produce entropy at such a faster rate. So a static
Universe is not entropically favoured in the presence of density
fluctuations.

On the other hand, the density perturbations no longer grow
exponentially in an expanding Universe. Instead they grow as a power
law because gravity has to work a lot harder against the Hubble drag
to form structures in such a Universe. The \autoref{eq:eleven} show
that the dissipation rate of the configuration entropy becomes smaller
in an expanding Universe. The configuration entropy rate depends on
the rate of expansion of the Universe and the growth rate of
perturbations inside it. Clearly, the configuration entropy rate would
depend on the cosmological model. In a matter dominated expanding
Universe, the configuration entropy rate becomes negative after a
large time. The Universe is slowed down and the growth rate of
structures is enhanced in such model. Consequently, the configuration
entropy would dissipate at a higher rate under such conditions. If the
other entropy generation processes are not as efficient as the
dissipation, it may again lead to an unphysical situation such as
$\frac{dS_{T}(t)}{dt}<0$. The validity of the second law of
thermodynamics demands that the dissipation rate of the configuration
entropy has to be always less than the collective entropy generation
rates from all the other sources.

The configuration entropy continues to dissipate in a matter dominated
Universe because the fluctuations on larger scales continue to grow.
The only way to control the dissipation is to suppress the growth of
structures by increasing the Hubble drag. This is exactly what we see
in a Universe dominated by the cosmological constant. In a lambda
dominated Universe, the configuration entropy converges to a constant
value after the growth of structures on linear scales are shut
off. The accelerated expansion of the Universe damps out the growth of
structures on linear scales and leads to an ideal situation where the
dissipation of the configuration entropy cease to exist. Here we would
like to mention that our analysis is only limited to the linear regime
and hence misses out the dissipation of the configuration entropy due
to the growth of non-linear structures.

The dissipation of the configuration entropy is also supported by the
fact that the information entropy decreases when a Gaussian
distribution turns in to a non-Gaussian distribution. The Gaussian
distribution has the maximum information entropy among all other
distributions with a specified variance. It is well known from the CMB
observations that the primordial fluctuations from the early Universe
were highly Gaussian \citep{planck14, planck16b} whereas the present
day mass distribution in the Universe is highly non-Gaussian. The
large scale structures in the Universe emerge from the growth of these
fluctuations via the process of gravitational instability which
developes non-Gaussianity in the distribution leading to a dissipation
of the information entropy.

Admittedly, the analysis presented in this work does not yet provide
an alternative to the dark energy. But it certainly points out to some
interesting links between the dark energy and the configuration
entropy of the Universe. Based on our observations, here we propose a
conjecture without providing a rigorous proof. The homogeneity and
isotropy is the most preferred and natural state of the Universe as it
corresponds to the state with maximum information. The fact that our
Universe is represented by the homogeneous and isotropic
Friedman-Lemaitre-Robertson-Walker (FLRW) metric on large scales is
possibly an evidence that the Universe maintains its status quo
despite the growth of perturbations in it. So the cosmological
principle is not merely a choice but a requirement. The Early Universe
was homogeneous and isotropic to a very high degree of precision and
thus possessed higher information entropy on all scales. The Universe
still retains maximum amount of information beyond the scale of
homogeneity and isotropy. But the mass distribution on small scales is
highly inhomogeneous and anisotropic due to the formation of
structures by gravitational clustering. Any deviation from homogeneity
and isotropy acts as a sink of information requiring a balance of
entropy. Each inhomogeneous and anisotropic patch of the Universe
below the scale of homogeneity and isotropy may thus experience an
entropic force resulting into an acceleration which ensures the
maximum entropy production principle. Besides the different
mass-energy densities in the Universe, the information entropy may
also take a decisive role in determining the large scale dynamics of
the Universe. It is interesting to note that if this is true then
there exists a natural explanation to the coincidence problem which
asks why the present densities of matter and dark energy happen to be
the same order of magnitude. If the accelerated expansion of the
Universe results from its response to the dissipation of the
configuration entropy in a matter dominated Universe then they are
expected to be of the same order of magnitude.
\section{ACKNOWLEDGEMENT}
I sincerely thank an anonymous reviewer for the useful comments and
suggestions. The author would like to acknowledge financial support
from the SERB, DST, Government of India through the project
EMR/2015/001037. The author would also like to acknowledge IUCAA, Pune
and CTS, IIT, Kharagpur for providing support through associateship
and visitors programme respectively.

\bsp	
\label{lastpage}
\end{document}